\documentclass[letterpaper,10pt]{article}

\usepackage{opticameet3} 

\newcommand\authormark[1]{\textsuperscript{#1}}

\usepackage{amsmath,amssymb}
\usepackage[colorlinks=true,bookmarks=false,citecolor=blue,urlcolor=blue]{hyperref} 
\usepackage{wrapfig}
\usepackage{subcaption}
\usepackage{url}

\definecolor{myred}{rgb}{1,0.2,0.2}
\definecolor{myblue}{rgb}{0,0.3,1}
\definecolor{mygreen}{rgb}{0.2,0.7,0}
\definecolor{myorange}{rgb}{1,0.5,0}

\begin{document}

\title{First Demonstration of 512-Color Shift Keying Signal Demodulation Using Neural Equalization for Optical Camera Communication}

\author{Yukito Onodera\authormark{1,*}, Daisuke Hisano\authormark{2}, Kazuki Maruta\authormark{3}, and Yu Nakayama\authormark{1}}
\address{
\authormark{1}Graduate School of Engineering, Tokyo University of Agriculture and Technology, 2--24--16 Nakacho, Koganei, Tokyo, 184--8588 Japan\\
\authormark{2}Graduate School of Engineering,
Osaka University,
2--1 Yamada-oka, Suita, Osaka 565--0871, Japan\\
\authormark{3}Faculty of Engineering, Tokyo University of Science, 6--8--1 Niijuku, Katsushika-ku, Tokyo, 125--8585 Japan
}
\email{\authormark{*}yukito.onodera@ynlb.org} 

\begin{abstract}
This paper experimentally demonstrates 512 color shift keying (CSK) signal transmission for optical camera communication (OCC). We achieved error-free operation with a CMOS image sensor module and a multi-label classification neural network-based equalizer.
\end{abstract}


\section{Introduction}
Optical Camera Communication (OCC) is a strong option for the next-generation optical wireless communication.
It leverages a complementary metal-oxide-semiconductor (CMOS) image sensor for a data receiver, and widespread commercial devices with embedded cameras can be employed as receiver devices.
OCC provides cost-efficient and license-free communication channels without using radio waves.
To enhance throughput has been one of the significant research topics for OCC, because the data rate is mainly limited by the exposure time and the frame rate of the receiver camera.
In addition, a flicker-free operation is an essential requirement.

\begin{figure}[!b]
\centering
  \subcaptionbox{CIE1931 color space~\cite{cie1931}. }{\includegraphics[width=0.3\textwidth]{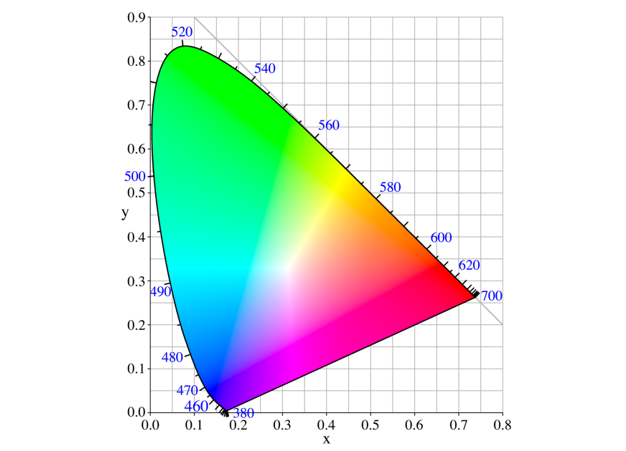}}
  \subcaptionbox{Spectral sensitivity~\cite{spectral}. }{\includegraphics[width=0.3\textwidth]{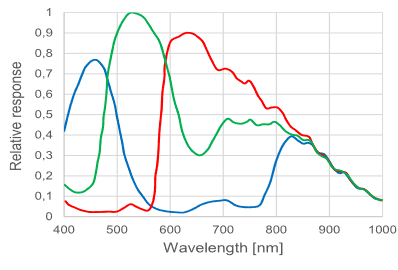}}
  \subcaptionbox{Related work. }{\includegraphics[width=0.3\textwidth]{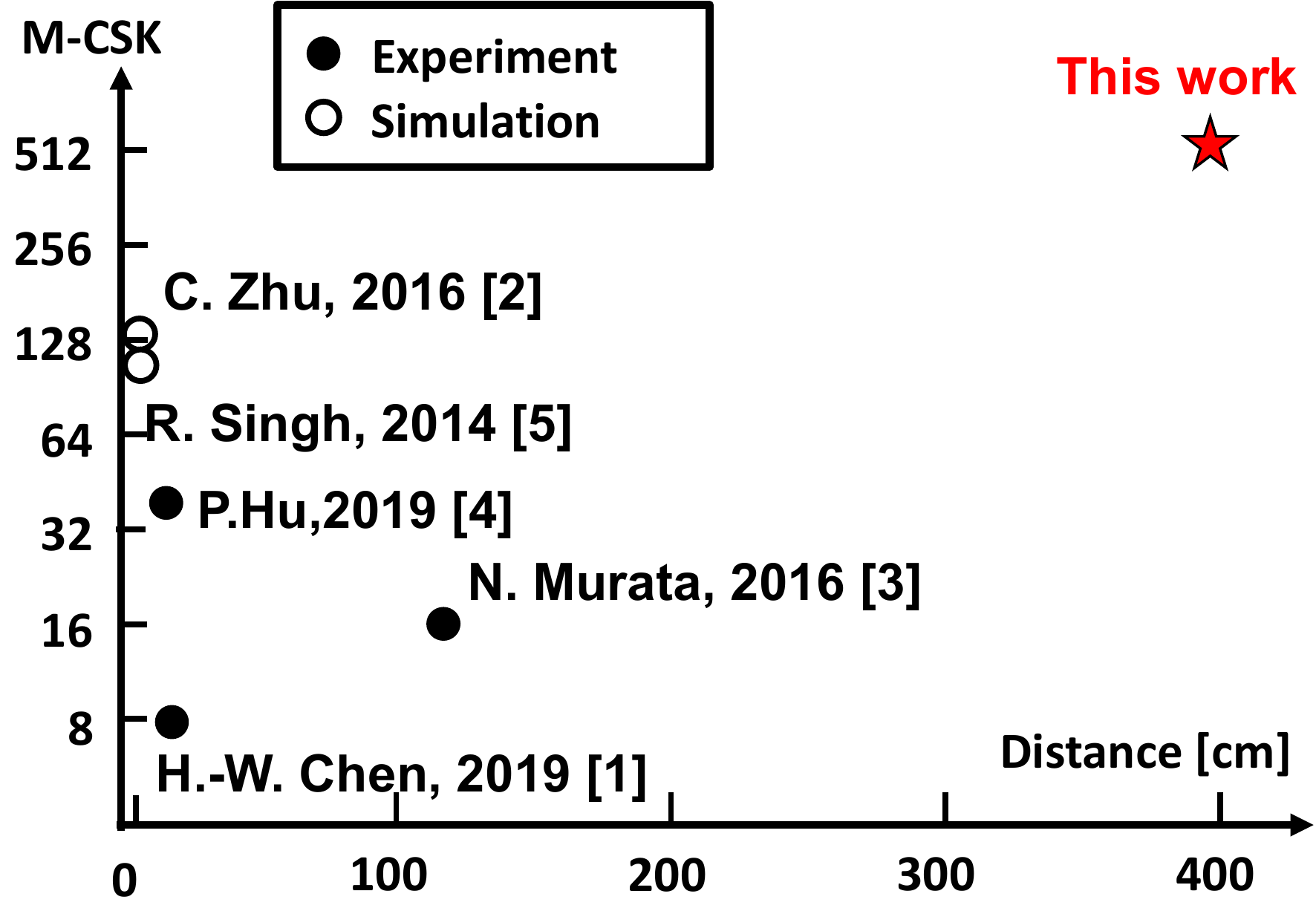}}
\vspace{-10pt}
\caption{Overview of OCC-CSK.}
\label{graph:intro}
\end{figure}

Color-Shift Keying (CSK) has attracted attention for increasing the data rate in OCC.
CSK is a VLC modulation scheme recommended by the IEEE 802.15.7 task group.
The color space chromaticity diagram defined by CIE 1931 (Fig.~\ref{graph:intro} (a)) is generally used for CSK.
CIE 1931 diagram maps all colors that are perceivable by human eyes to the $(x, y)$ color space.
It is then converted to the emission intensity of the RGB-LED transmitter.
However, an optical camera generally has spectral sensitivity characteristics as exemplified in Fig.~\ref{graph:intro} (b) and RGB components are extracted through color filters.
This nature causes crosstalk among colors which should be canceled via digital signal processing.
As related works, 8-CSK data transmission over $4$ cm was presented in \cite{chen2019color}.
16-CSK over $80$ cm using a quadrichromatic LED was reported in \cite{liang2017constellation}.
In \cite{murata2016digital}, 16-digital CSK over $100$ cm was achieved based on IEEE 802.15.7 CSK constellations.
Tri-LEDs based 32-CSK over $3$ cm was demonstrated in \cite{hu2015colorbars, hu2019high}.
More high-modulation schemes have only been investigated in theory and computer simulation~\cite{singh2014enhanced, zhu2016hierarchical}.
The above recent achievements are summarized in Fig.~\ref{graph:intro} (c).

This paper presents the first demonstration of $512$-CSK signal error-free demodulation using Sony's IMX530 CMOS image sensor and $50$-mm optical lens.
The nonlinear crosstalk is compensated with neural equalization from the received CMOS image sensor raw data.

\section{Receiver Configuration}
\begin{figure}[t]
\centering
	\includegraphics[width=\textwidth]{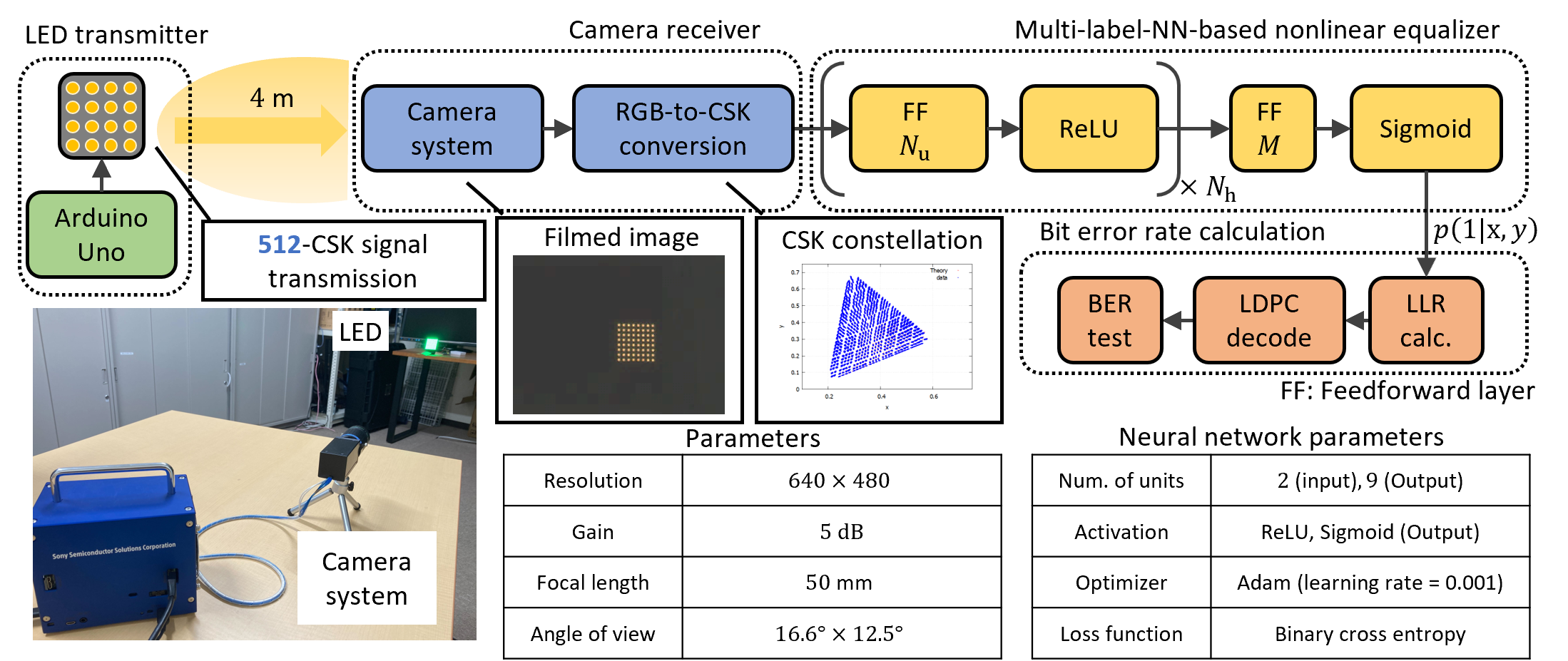}
	\caption{Receiver configuration and experiment setup.}
	\label{fig:receiver}
\end{figure}

Fig.~\ref{fig:receiver} shows the receiver block diagram.
We employed the camera system provided by Sony Semiconductor Solutions Corporations as a receiver which has a 12-bit resolution for respective RGB sensitivity.
It can also output purely raw image data without demosaicing, denoising, and white balancing.
An optical signal is transmitted through 8$\times$8 LED planar array and received via $50$ mm optical lens.
The raw data of the image sensor are converted from RGB color space to CIE 1931 format for CSK demodulation using the following formula;
\begin{equation}
    \left(
        \begin{array}{c}
            x \\
            y
        \end{array}
    \right)
    =
    \left(
        \begin{array}{ccc}
            0.4124 & 0.3576 & 0.1805 \\
            0.2126 & 0.7152 & 0.0722
        \end{array}
    \right)
    \left(
        \begin{array}{c}
            R \\
            G \\
            B
        \end{array}
    \right).
\end{equation}
They are then transferred to the multi-label neural network (NN)-based equalizer to erase the nonlinear crosstalk between color channels. 
The numbers of input units, hidden units, and output units are $2 (=x,y)$, $N_u$, $M$, respectively.
$M$ corresponds to the number of bits per symbol.
In this paper, we employ 512-CSK, so $M = \log_{2} 512=9$ units.
The number of hidden layers is $N_h$.
Log likelihood ratio (LLR) is calculated from the posterior probability distribution $p(1|x,y)$ obtained with the multi-label NN~\cite{Shental2019DNN}.
Finally, LLR is input to the low-density parity-check (LDPC) decoder.
Symbols for 512-CSK are sequentially placed with bits in a triangular manner, starting from the vertex of blue ($x = 0.1805, y = 0.0722$).

\section{Experiment Results}
Specifications of the camera and lens are described in Fig.~\ref{fig:receiver}.
We used $8 \times 8$ LED planar array and its panel size is $6.5$ cm.
The number of LEDs was varied from 1$\times$1 to 8$\times$8 in order to evaluate BER characteristics based on the area occupied by LEDs in the captured image (light intensity).
The transmission distance was $4$ m.
The experiments were conducted in a dark room to avoid contamination by external light.
The main parameters for NN are also listed in Fig.~\ref{fig:receiver}.
The numbers of hidden layers and units are evaluation variables to be optimized.
15000 samples were used for NN training with 5000 epochs.
The batch size was set to 4096.
Parameters for LDPC conforms to Digital Video Broadcasting--Satellite2 (DVB-S.2).
The codeword length is 64800 bits per block and 3 blocks are transmitted for BER calculation.
BER performance of 512-CSK is evaluated using various code rates from 1/4 to 9/10.

\begin{figure}[t]
\centering
	\includegraphics[width=\textwidth]{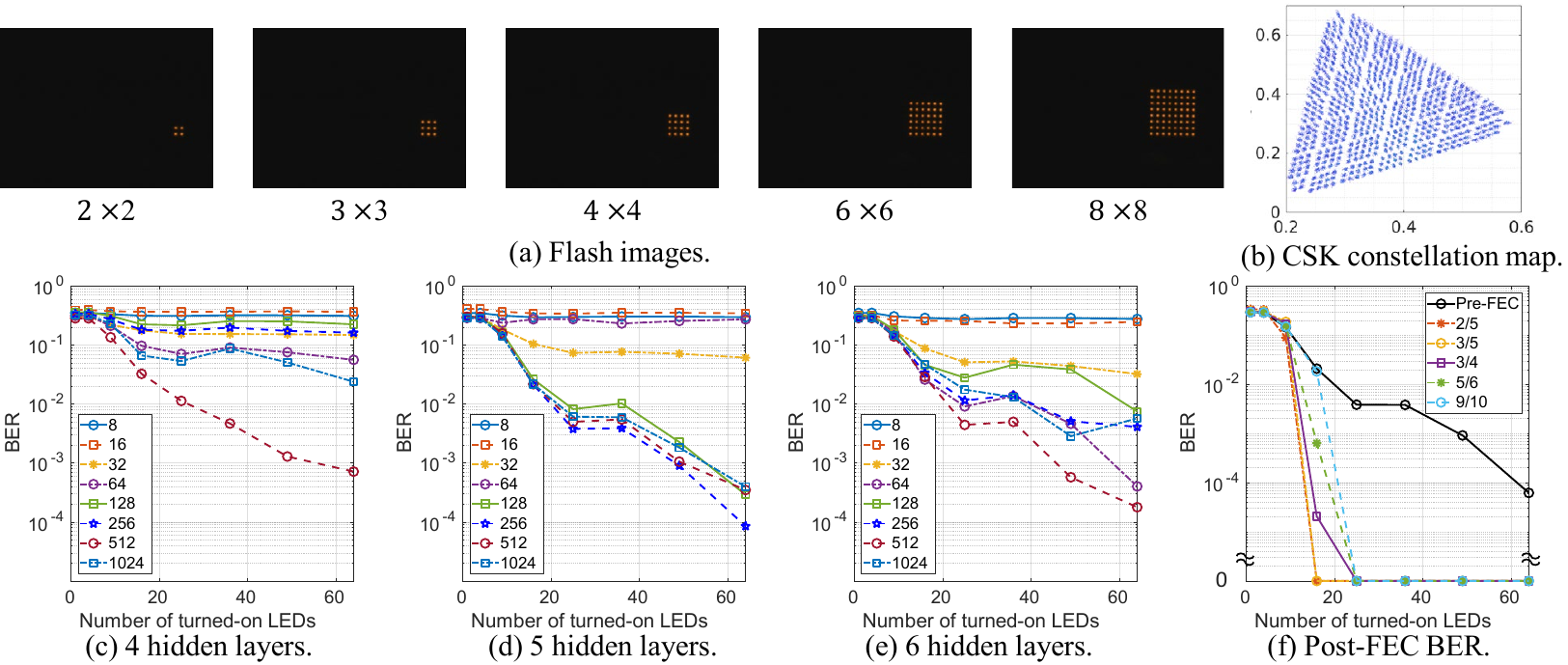}
	\caption{Experiment results.}
	\label{fig:exp_res}
\end{figure}

Fig.~\ref{fig:exp_res} summarizes the measured BER performance with the number of turned-on LEDs for 512-CSK.
Examples of flash images for LED array from 2$\times$2 to 8$\times$8 are shown in Fig.~\ref{fig:exp_res} (a).
It can be confirmed that the reception accuracy would vary on the number of light sources.
Received constellation map for 512-CSK is visualized in Fig.~\ref{fig:exp_res} (b).
The slight cracks can be observed due to the quantization level; 512-CSK symbols is expressed by 100 steps in each RGB light emission and the receiver with 12 quantization bits can recognize such fine-grained signals.
Figs.~\ref{fig:exp_res} (c)--(e) plots uncoded BER performances with the number of turned-on LEDs to optimize NN parameters.
Legends in each figure indicate the number of units.
NN structured by 256 units and 5 hidden layers can achieve the lowest BER performance with avoiding overfitting.
Finally, Fig.~\ref{fig:exp_res} (f) presents LDPC-coded BER with the number of turned-on LEDs based on the above optimized NN parameters.
If the receiver can capture 5$\times$5 LEDs, sufficiently low BER is attained even with a higher coding rate overheaded by up to 10\%.
It should be noted that we transmitted about 2$\times$10$^{6}$ bits which guarantees BER below 10$^{-6}$ even though the results show BER$=$ 0.
As a result, we have experimentally verified that our proposed scheme significantly outperforms existing CSK in terms of modulation order as well as transmission distance.


\section{Conclusion}
This paper demonstrated the first experimental results of $512$-CSK for OCC.
We achieved error-free transmission over $4$ meters using high-resolution raw data from Sony's IMX530 CMOS image sensor, neural network based equalization, and LDPC forward error correction.
This work significantly contributes to enhancing the OCC data rate for next-generation visible light communication.

\section*{Acknowledgment}
A part of this work was supported by Sony Semiconductor Solutions Corporation, JST Presto Grant Number JPMJPR2137, and JST START Project Promotion Type (Supporting Small Business Innovation Research (SBIR) Phase 1), Grant Number JPMJST2260, Japan.

\bibliographystyle{opticajnl.bst}
\bibliography{bibliography}


\end{document}